\documentclass[aps,pre,twocolumn,showpacs,groupedaddress,superscriptaddress,longbibliography]{revtex4-1}

\usepackage{natbib,hyperref}
\hypersetup{colorlinks=true, citecolor=blue, urlcolor=blue, linkcolor=blue}
\usepackage[english]{babel}
\usepackage[utf8]{inputenc}
\usepackage{graphicx}
\usepackage[caption=false]{subfig}
\usepackage{amsmath}
\usepackage{amsfonts}
\usepackage{amssymb}
\begin{document}


\title{Dimensional coupling induced current reversal in two-dimensional driven lattices}
\author{Aritra K. Mukhopadhyay}
\email{Aritra.Mukhopadhyay@physnet.uni-hamburg.de}
\affiliation{Zentrum f\"ur Optische Quantentechnologien, Universit\"at Hamburg, Luruper Chaussee 149, 22761 Hamburg, Germany}
\author{Tianting Xie}
\email{xietianting@hotmail.com}
\affiliation{College of Mathematics, Sichuan University, Chengdu, 610065, China}
\author{Benno Liebchen}
\email{liebchen@hhu.de}
\affiliation{SUPA, School of Physics and Astronomy, University of 
	Edinburgh, Peter Guthrie Tait Road, Edinburgh, EH9 3FD, UK}
\affiliation{Institute for Theoretical Physics II: Soft Matter, Heinrich-Heine Universit\"at D\"usseldorf, Universit\"atsstrasse 1, 40225, D\"usseldorf, Germany} 
\author{Peter Schmelcher}
\email{Peter.Schmelcher@physnet.uni-hamburg.de}
\affiliation{Zentrum f\"ur Optische Quantentechnologien, Universit\"at Hamburg, Luruper Chaussee 149, 22761 Hamburg, Germany}
\affiliation{The Hamburg Centre for Ultrafast Imaging, Universit\"at Hamburg, Luruper Chaussee 149, 22761 Hamburg, Germany}

\date{\today}

\begin{abstract}
 We show that the direction of directed particle transport in a two dimensional ac-driven lattice can be dynamically reversed by changing the structure of the lattice in the direction perpendicular to the applied driving force. These structural changes introduce dimensional coupling effects, the strength of which governs the timescale of the current reversals. The underlying mechanism is based on the fact that dimensional coupling allows the particles to explore regions of phase space which are inaccessible otherwise. The experimental realization for cold atoms in ac-driven optical lattices is discussed.
\end{abstract}


\maketitle

\section{Introduction} 
The ratchet effect allows to create a directed particle transport in an unbiased non-equilibrium environment and thus to extract mechanical  
work from a fluctuating bath \cite{Hanggi2005,Astumian2002,Hanggi2009}. Such a conversion is impossible for macroscopic equilibrium systems and makes the ratchet effect a fundamental 
non-equilibrium phenomenon.
While originally conceived as proof-of-principle examples 
of rectification schemes producing work from fluctuations \cite{Prost1994,Magnasco1993,Bartussek1994,Faucheux1995} and 
as possible explanations for the mechanism allowing molecular motors to show directed motion along cytoskeleton filaments 
\cite{Astumian1997,Julicher1997}, ratchets now form a widespread paradigm with a large realm of applications in atomic, condensed matter and biophysics. 

Specific applications range from the rectification of atomic \cite{Schiavoni2003a}, colloidal \cite{Rousselet1994} and 
bacterial motion \cite{DiLeonardo2010,Koumakis2013,Reichhardt2017,Vizsnyiczai2017} to the transportation of 
fluxons in Josephson junctions arrays \cite{Falo1999,Zolotaryuk2012} and vortices in conformal crystal arrays \cite{Reichhardt2016,Reichhardt2015}.
Very recently, it has been demonstrated, that ratchets also allow  
to control the dynamics of topological solitons in ionic crystals \cite{Brox2017}, enhance photocurrents in quantum wells \cite{Faltermeier2017}, can rectify the chirality of magnetization in artificial spin ice \cite{Gliga2017} and create a light modulated electron transport across organic bulk 
heterojunctions \cite{Kedem2017}.

\begin{figure*}[t]
	\includegraphics[scale=0.048]{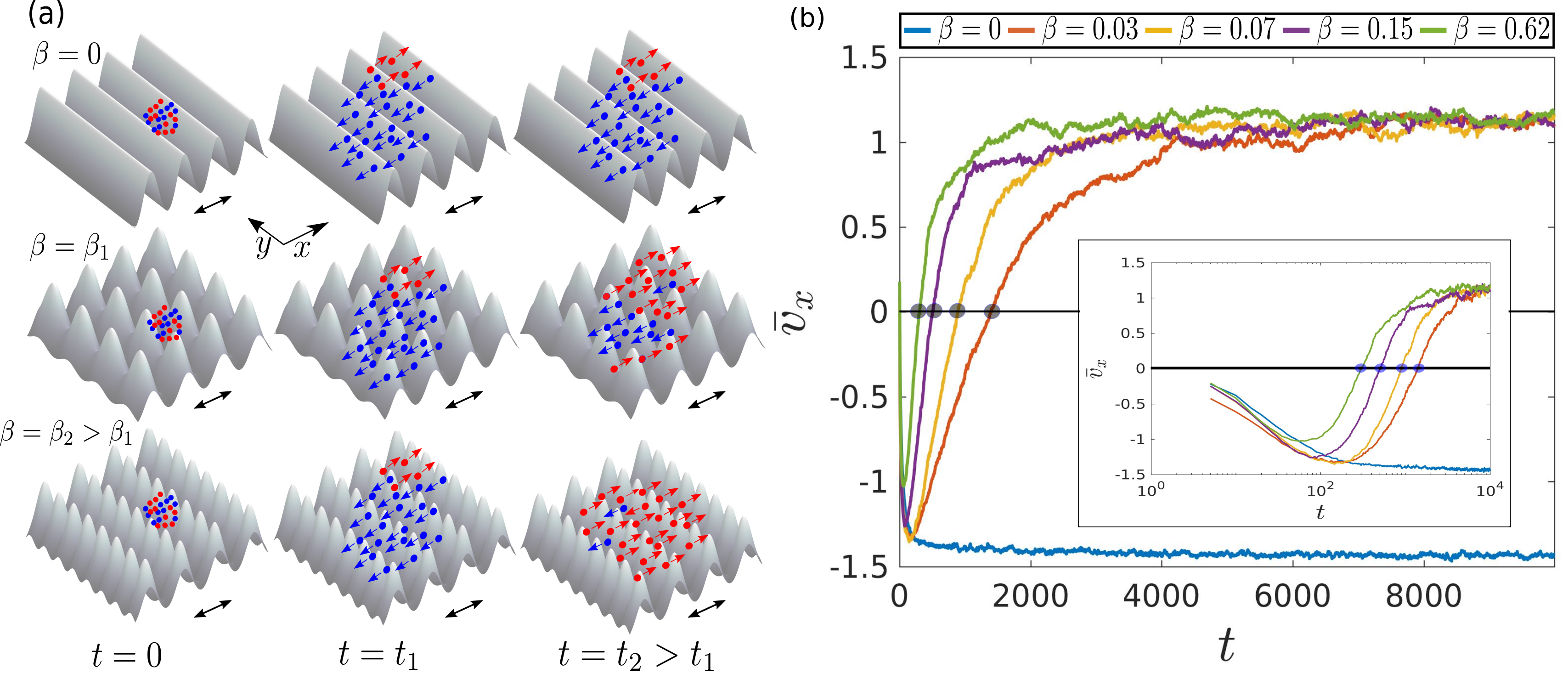}
	\caption{(a) Schematic diagram of the setup demonstrating the phenomenon of dimensional coupling induced current reversal. The filled dots denote particles and the color (red/blue) indicate the sign of the $x$ component of their velocities (right/left). In a driven quasi-1D lattice (upper panel), most particles travel in the negative $x$-direction resulting in an average transport in this direction. However in the driven 2D lattice (middle and lower panel) having non-zero dimensional coupling $\beta$, most particles initially travel towards the negative $x$-direction but at later times revert their movement, resulting in a dynamical current reversal. Larger values of the coupling $\beta$ reduces the timescale of the current reversal (lower panel). (b) Mean transport velocity of the ensemble along the $x$-direction as a function of time for different values of $\beta$ for a linear and logarithmic (inset) timescale. The grey circles correspond to the reversal timescales $t_{r,\beta}$ for different values of $\beta$. Remaining parameters: $U=0.88$, $a=0.48$, $\alpha=9.61$ and $\gamma=0.62$. There is no directed transport of particles along the $y$-direction.}\label{fig1}
\end{figure*}

While the fact that a specific setup creates a directed particle transport can typically be predicted based on symmetry properties
\cite{Flach2000,Denisov2014}, the strength and even the direction of the emerging currents are far less immediate. 
In fact, the current direction can often be reverted by changing the value of a certain control parameter or the properties 
of the rectified objects (e.g. their mass or mobility), without changing the symmetry of the 
underlying equations. Achieving such current reversals is the key aim of many investigations, as they allow segregation of particle mixtures by transporting 
particles of e.g. different mass or mobility in opposite directions, where they can be collected. 

While most ratchets are still studied in one spatial dimension (1D) \cite{Reimann2002,Hanggi2009}, particularly those 
operating in the Hamiltonian regime \cite{Flach2000,Schanz2001,Schanz2005,Denisov2014,Wulf2012,Liebchen2012}, 
recent experiments have significantly progressed regarding the 
construction of highly controllable two-dimensional (2D) ratchet devices. These include 
cold atoms in ac-driven optical lattices \cite{Lebedev2009,Cubero2012,Renzoni2009} and the very recent example of a fully
configurable 2D ratchet based on colloids in holographic optical tweezers \cite{Arzola2017}.
Conceptually, the key new ingredient in 2D ratchets is the coupling between the dimensions, which has been shown to allow, in the overdamped regime, for a
directed transport at an angle relative to the driving law \cite{Arzola2017, Mukhopadhyay2017} and may 
also involve transportation completely orthogonal to the driving \cite{Reichhardt2003}.
In the present work, we demonstrate that dimensional coupling can even lead to current reversals. 

A 2D potential landscape having a periodic potential along, for e.g., the $x$-direction but without any potential variation along the perpendicular $y$-direction (henceforth referred to as `quasi-1D lattice') allows for directed particle transport when driven by an appropriately chosen ac-driving force in the $x$-direction (see Fig.~\ref{fig1}a upper panel). Keeping the driving unchanged but performing a structural change of the lattice along the $y$-direction introduces dimensional coupling effects. We show that this coupling does not affect the directed particle current for short timescales, but reverts its direction at longer timescales as compared to the quasi-1D lattice (see Fig.~\ref{fig1}a lower panel). These dimensional coupling induced current reversals (DCIR) occur dynamically in time \cite{Liebchen2012}, as opposed to the standard scenario of asymptotic current-reversals due to a change of system parameter where the direction of current is
time-independent \cite{Marconi2007,Mateos2000,DeSouzaSilva2006}.
We show that the reversal timescale can be varied by thousands of driving period by varying the structure of the lattice perpendicular to the driving direction (see Fig.~\ref{fig1}a middle panel). The underlying mechanism of these current reversals uses the fact that changing the structure of the lattice along the second dimension allows the particles to explore different regions of phase space which are inaccessible in the quasi-1D lattice.

\section{Setup} 
We consider $N$ non-interacting classical particles in a 2D lattice 
of elliptic Gaussian barriers laterally driven along the $x$-direction via an external bi-harmonic driving force 
$f(t)= d_{x} (\sin \omega t + 0.25\sin (2\omega t + \pi/2))$. Here, $d_x$ and $\omega$ 
are the amplitude and the frequency of the driving, thereby introducing a temporal periodicity of $T=2\pi / \omega$. 
The system is described by the Hamiltonian: 
\begin{eqnarray}
&H&=\frac{p_x^2}{2m}+\frac{p_y^2}{2m} \nonumber\\
&+&\sum_{i,j=-\infty}^{+\infty} V e^{-\left[\beta_x\left(x-f(t)-(i+\frac{1}{2})L_x\right)^2+\beta_y\left(y-(j+\frac{1}{2})L_y\right)^2\right]}\label{hamil}
\end{eqnarray}
where the potential barriers have a height $V$ and the equilibrium distances between them 
along $x$ and $y$ are given by $L_x$ and $L_y$ respectively. 
This potential breaks both the parity $ x\rightarrow - x + \chi$ symmetry along the $x$ direction and the 
time-reversal $t\rightarrow -t +\tau$ symmetry (for all possible constants $\chi$ and $\tau$), 
while preserving parity symmetry along the $y$ direction. Possible realizations of this setup include cold atoms in optical lattices, at microkelvin temperatures, where a classical description is appropriate \cite{Renzoni2009} and which to a good approximation represents a Hamiltonian setup.

Introducing dimensionless variables $x'=\frac{x}{L_x}$, $y'=\frac{y}{L_y}$ and $t'=\omega t$ 
and dropping the primes for simplicity, 
the equation of motion for a single particle at position  ${\bf r}$ with momentum $\bf p$ reads
\begin{equation}
\ddot{\bf r} = \sum_{m,n=-\infty}^{+\infty} \mathcal{U}\left( {\bf r} - F(t) {\bf e_x} - {\bf R}_{m,n} \right) e^{-\mathcal{G}({\bf r} - F(t) {\bf e_x} - {\bf R}_{m,n})}\label{eqm2}
\end{equation}
where $F(t)=\left( a\sin t + 0.25a\sin (2 t + \pi/2) ,0\right)$ 
is the effective driving law, ${\bf e_x}=(1,0)$, ${\bf R}_{m,n}=(m,n)$ 
denotes the positions of the maxima of the Gaussian barriers
where $\left(m-\frac{1}{2}\right)$,$\left(n-\frac{1}{2}\right)$ $\in\mathbb{Z} $ and $\mathcal{U}({\bf r})=\left(Ux,\beta Uy \right)$, $\mathcal{G}({\bf r})=\alpha (x^2 + \gamma y^2)$. 
The parameter space of our system is therefore essentially five-dimensional, where the 
dimensionless parameters are given by 
a reduced barrier height $U=\frac{2V\beta_x}{m\omega ^2}$, an effective driving amplitude $a=\frac{d_x}{L_x}$, 
as well as the two parameters, 
$\alpha=\beta_x L_x^2$ and $\gamma=\frac{\beta_y L_y^2}{\beta_x L_x^2}$,
characterizing the localization of the 
Gaussian barriers along the $x$ and $y$ directions. A final key control parameter is 
$\beta=\frac{\beta_y}{\beta_x}$ which measures the coupling between the two dimensions. 
The limits $\beta\rightarrow 0$ and $\beta\rightarrow \infty$ both correspond to quasi one 
dimensional lattices. 

\section{Results}
\begin{figure}[t]
	\includegraphics[scale=0.11]{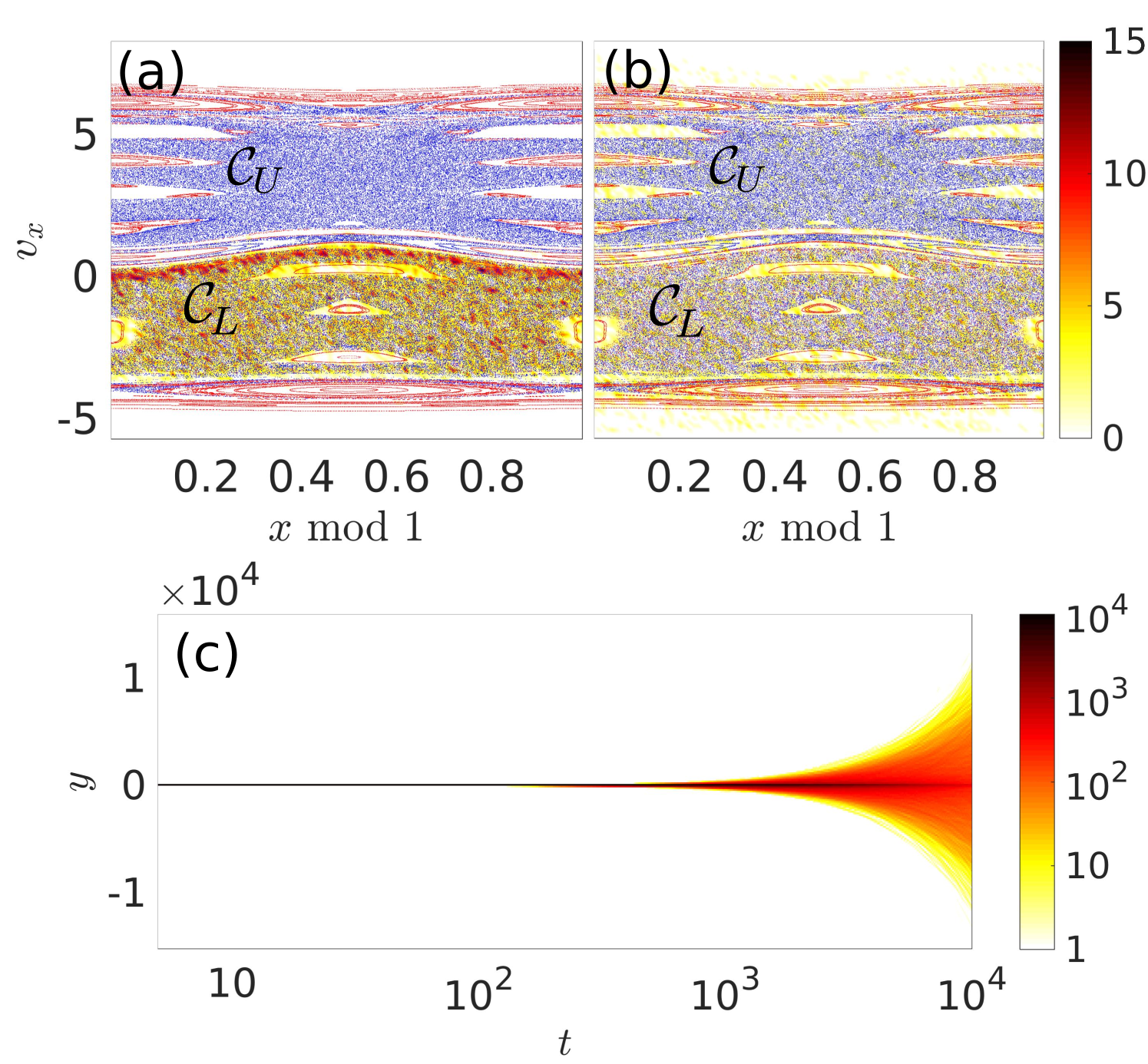}
	\caption{The particle distribution as a function of position $x$ mod $1$ and $v_x$ (in colormap) of all the $N=10^4$ particles  propagating in the 2D lattice with $\beta=0.03$ superimposed on the PSOS of the quasi 1D driven lattice (regular islands in red and chaotic seas in blue) at (a) $t=50$ and (b) $t= 1.5\times 10^3$. (c) The particle distribution as a function of $y$ and $t$ showing the spreading of the ensemble along the $y$-direction with time.}
	\label{fig2}
\end{figure}
To explore the transport properties of our setup, 
we initialize $N=10^4$ particles with small random velocities $v_x,v_y \in [-0.1,0.1]$ such that their 
initial kinetic energies are small compared to the potential height of the lattice. 
In order to mimic a localized loading of particles into the lattice, we initialize the particles within the square regions $[-0.1,0.1]\times [-0.1,0.1]$ 
centered around the potential minima of the lattice. Subsequently we time evolve our ensemble up to $t = 10^4$ by numerical integration of Eq.~\ref{eqm2} 
using a Runge-Kutta Dormand Prince integrator.

For $\beta=0$, the lattice is quasi-1D (upper panels in Fig. 1a) and produces a non-zero mean velocity 
pointing in negative $x$-direction (Fig.~\ref{fig1}b). This behaviour is expected since the system breaks both the parity and time reversal symmetry along the $x$ direction, thus satisfying the necessary criteria for a non-zero directed transport \cite{Flach2000,Schanz2005,Schanz2001}. Since there is no driving in the $y$-direction, the symmetries are preserved and hence there is no directed transport along this direction. The transport in the $x$-direction accelerates until it finally saturates at $\bar{v}_x\simeq-1.4$. 

We now vary $\beta$ to explore the impact of dimensional coupling effects on the directed transport. 
As shown in Fig.~\ref{fig1}b, for $\beta=0.03$, 
the early time transport velocity is negative and approaches a similar speed of $\bar{v}_x\simeq-1.35$, as in the quasi-1D case at around $t\simeq 1.5\times 10^2$.
Remarkably, thereafter the transport begins to slow down and vanishes at $t=t_{r,\beta=0.03}\simeq 1.4 \times 10^3$.
Further on, it changes sign which leads to a current reversal. 
Finally, it approaches an asymptotic constant value of $\bar{v}_x\simeq1.2$.
Therefore, the structural change of the lattice in the direction orthogonal to the driving force reverts the transport direction. 

To study this dimensionality-induced current reversal in more detail, we perform our simulations for a stronger dimensional coupling $\beta=0.15$ and $\beta=0.62$, which leads to a qualitatively similar behaviour (see Fig.~\ref{fig1}b).
However, we find that the timescale at which the reversal occurs strongly depends on the strength of the dimensional coupling coefficient $\beta$.
Specifically for $\beta=0.62$, we obtain $t_r\simeq 3\times 10^2$ showing that the reversal timescale can be tuned by at least a factor of five.

\section{Discussion}

The underlying mechanism of the DCIR effect depends on two generic ingredients: 
(i) the existence of a mixed phase space (containing regular and at least two disconnected chaotic components) in the underlying quasi 1D lattice and (ii) 
the diffusional spreading dynamics in the 2D lattice along the orthogonal direction. We now discuss the occurrence of negative transport 
in the quasi 1D lattice ($\beta=0$) and will then analyze how the dimensional coupling effect can revert the transport direction.

Due to the absence of forces acting along the $y$-direction, the dynamics in the 
quasi 1D lattice (Fig.\ref{fig1}a) can be decomposed into a constant drift in $y$-direction and a motion in a 1D lattice driven along the $x$-axis.
The latter case is described by a three-dimensional (3D) phase space illustrated by taking stroboscopic snapshots of 
$x(t), v_x(t)$ at $t=n (n\in \mathbb{N})$ of particles with different initial conditions.
This leads to Poinc\'{a}re surfaces of section (PSOS) as shown in Fig.~\ref{fig2}a where the reflection symmetry about $v_x=0$ is broken. This PSOS is characterized by two prominent chaotic components or `seas': the upper sea $\mathcal{C}_U$ 
between $0.75\lesssim v_x \lesssim 6.0$ and the lower sea $\mathcal{C}_L$ between $-3.5\lesssim v_x \lesssim 0.2$. 
These chaotic seas are separated from each other by regular invariant spanning curves at $v_x \simeq 0.2$ preventing particles 
to travel between the chaotic components. Hence particles initialized with low initial energies $v_x \in [-0.1,0.1]$ and  
occupying $\mathcal{C}_L$, matching the initial conditions used in our simulations, undergo chaotic diffusion through the lattice 
with negative velocities along the $x$-direction until they are uniformly distributed over $\mathcal{C}_L$. As a result, 
we observe a negative directed transport of the ensemble.

Let us now explore the mechanism allowing dimensional coupling ($\beta > 0$) to revert the transport direction: 
In this case, the phase space is five-dimensional (5D) characterized by $(x,v_x,y,v_y,t)$ which complicates both the illustration and analysis of the transport based on the phase space structures. However up to a certain timescale, the dynamics of the particles even in 
this higher dimensional phase space can be effectively understood in terms of the dynamic occupation of the ensemble in 
the quasi 1D PSOS. To show this, we superpose the snapshots of the ensemble particle coordinates $(x,v_x)$ for $\beta=0.03$ on 
the quasi 1D PSOS at two different times $t=50$ and $t=1.5\times 10^3$ (Fig.~\ref{fig2}). At $t=50$, well before 
the reversal timescale $t_{r,\beta=0.03}=1.24\times 10^3$, the ensemble population is confined to $\mathcal{C}_L$ in a similar way as we have observed for 
$\beta=0$ (Fig.~\ref{fig2}a).
Physically, this results from the fact that at shorter timescales the particles experience comparatively 
strong driving forces which allow them to quickly move along the $x$-direction while in $y$-direction they move only very slowly with  a velocity largely dictated by the
initial conditions. Therefore, for a long time, they stay close to the 
potential valleys at $y=0$ (Fig.~\ref{fig2}c) where they hardly experience the 2D landscape of the potential.
 
 \begin{figure}[t]
 	\includegraphics[scale=0.10]{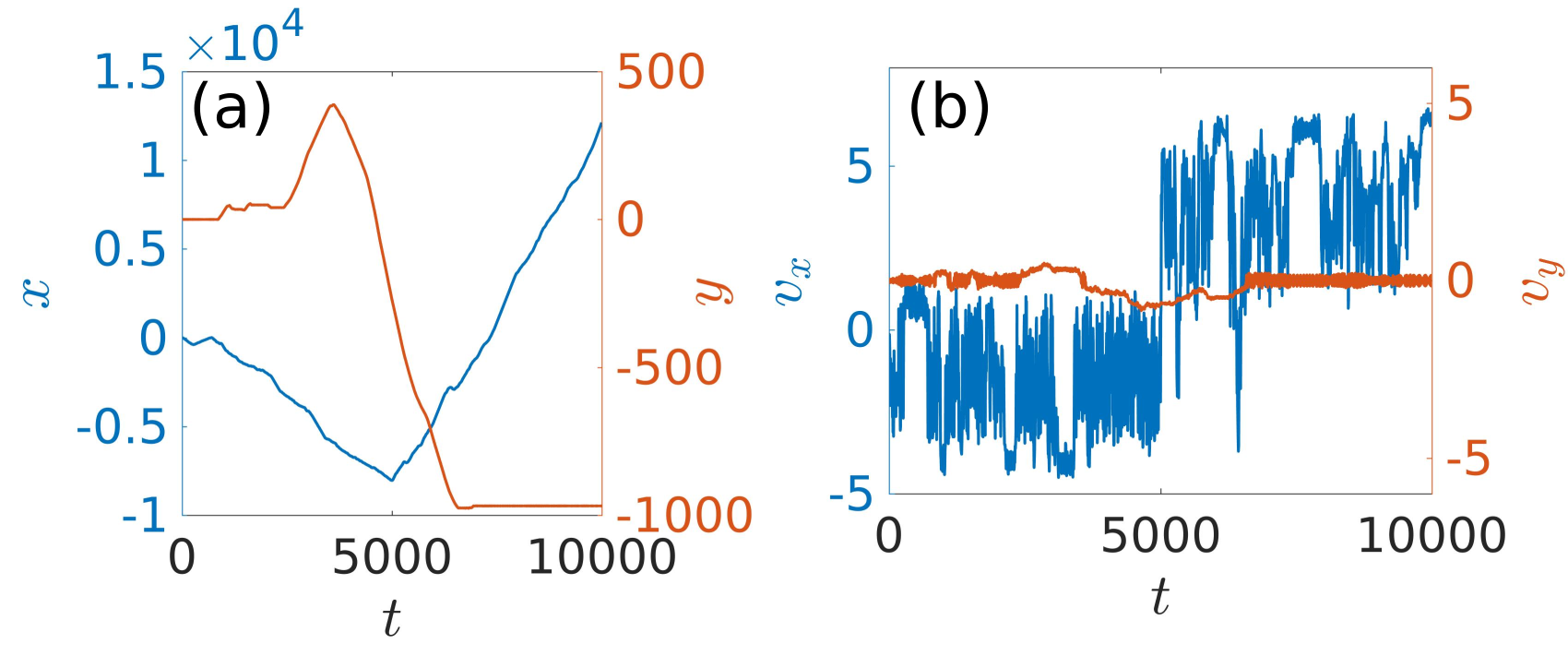}
 	\caption{The time dependence of (a) position and (b) velocity of a typical particle in the 2D lattice with $\beta=0.03$ initialized in the lower chaotic sea $\mathcal{C}_L$ (Fig.\ref{fig2}a) demonstrating the crossover to the upper chaotic sea $\mathcal{C}_U$ (Fig.\ref{fig2}b). Remaining parameters are the same as in Fig.\ref{fig1}b. Note that for this particular trajectory, the crossover happens at $t\simeq 5\times 10^3$, which is larger than the average reversal timescale $t_{r,\beta=0.03}=1.24\times 10^3$ of the ensemble.}
 	\label{fig3}
 \end{figure}

As time evolves, particles experience more and more of the 2D character of the potential which effectively transfers motion in $x$-direction 
into motion along the $y$-direction leading to a symmetric spreading of the ensemble along the $y$-direction (Fig.~\ref{fig2}c) . 
Particles are therefore no longer dictated by the structure of the 1D phase space but 
can explore the entire 5D phase space. 
They can, in particular, now freely cross the invariant spanning curves at $v_x\simeq 0.2$ of the 1D phase space to attain significant
positive velocities (Fig.~\ref{fig2}b). During the phase of temporal evolution when the particles can cross the invariant curve, the directed current slows down and reduces to zero. It finally becomes positive, since the asymptotic average velocity of the particles along the positive $x$-direction is higher than that along the negative $x$-direction. A typical trajectory demonstrating the crossover from $\mathcal{C}_L$ to $\mathcal{C}_U$ is shown in Fig.~\ref{fig3}.

\section{Control of the current reversal}

\begin{figure}[b]
	\includegraphics[scale=0.09]{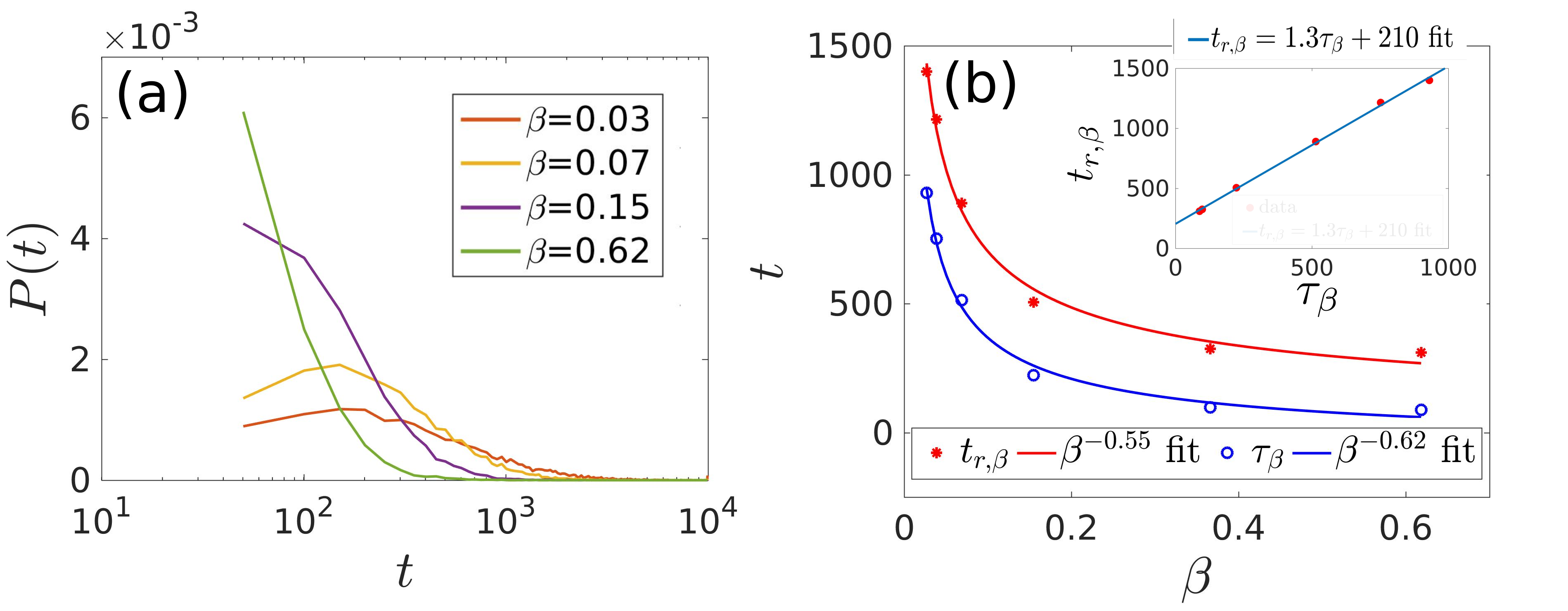}
	\caption{(a) The probability density $P(t)$ of the first crossing time (FCT) $t$ required by a particle to cross one lattice site along the $y$-direction for the first time. (b) The mean FCT $\tau_{\beta}$ (in blue) and the reversal timescale $t_{r,\beta}$ (in red) as functions of $\beta$ with corresponding inverse power law fits. The inset shows the linear relationship between $t_{r,\beta}$ and $\tau_{\beta}$.}
	\label{fig4}
\end{figure}

Let us finally discuss the dependence of the current reversal time $t_{r,\beta}$ on the parameter $\beta$. Following the above-outlined physical picture, the current reversal occurs at time scales comparable to the time a particle needs to experience a significant deviation from the neighborhood of the minimum of the lattice potential along the $y$-direction. For a particular value of $\beta$ and a given set of initial conditions, one can thus expect the reversal timescale $t_{r,\beta}$ to depend linearly on the average time $\tau_{\beta}$ the particles need to cross one lattice site along the $y$-direction for the very first time. In order to estimate $\tau_{\beta}$ for different values of $\beta$, we simulate ensembles of $10^4$ particles each with initial conditions identical to that used in our setup (Fig.~\ref{fig1}), but for different $\beta$ values and calculate the corresponding probability density $P(t)$ of the first crossing time (FCT) $t$ required by a particle to cross one lattice site along the $y$-direction (Fig.~\ref{fig4}a). As $\beta$ increases, the particles are likely to have shorter FCT and hence can experience the 2D landscape of the potential much earlier. This can be clearly seen in the Fig.~\ref{fig4}b (blue) which shows that the mean FCT $\tau_{\beta}$ decreases with increasing $\beta$ following a $\tau_{\beta} \sim \beta^{-0.6}$ power law. Confirming our expectation, a linear fit is shown to describe the relation between $t_{r,\beta}$ and $\tau_{\beta}$ to a good approximation (see Fig.~\ref{fig4}b (inset)) and hence $t_{r,\beta}$ follows a similar inverse power law $t_{r,\beta}\sim \beta^{-0.55}$ (Fig.~\ref{fig4}b, red). The reversal timescale depends also (weakly) on the initial velocities of the particles and we verified that a decrease of the initial velocity by a factor of 0.01 increases the reversal timescale approximately by a factor of 1.34.
 
\section{Experimental realization}
A setup to experimentally observe dimensional coupling-induced current reversals are cold atoms in optical lattices generated by laser beams in the regime of $\mu K$ temperatures where a classical description is appropriate \cite{Renzoni2009}. Setups based on holographic trapping of atoms \cite{Nogrette2014,Kim2016,Barredo2016,Stuart2018} might also provide an interesting and highly controllable alternative.
The resulting lattice can be driven by phase modulation using acousto-optical modulators and radio frequency generators. 
Translating our parameters to experimentally relevant quantities for an optical lattice 
setup with cold rubidium (Rb$^{87}$) atoms and $780\ nm$ lasers, we obtain the lattice height 
$V \sim 22 E_r$, the width $\frac{1}{\sqrt{\beta_x}}\sim 252\ nm$, the driving frequency $\omega \sim 10\omega_r$ 
and the driving amplitude $d_x \sim 390\ nm$, where $E_r$ and $\omega_r$ are the recoil energy and recoil frequency of the atom respectively. 
Interaction , disorder and noise effects \cite{Liebchen2012,Liebchen2015,Wulf2014}, 
would probably lead to a slow accumulation of particles within the regular portions of the phase space which may also aid 
them in crossing the regular barrier confining the initial conditions in the quasi 2D case to negative and only weakly 
positive velocities and may therefore lead to a slight decrease of the reversal time. 

\section{Concluding remarks} 
Dimensional coupling effects in two-dimensional lattices create a new route to produce current-reversals. 
Conversely to most other cases, the current reversal occurs dynamically here with a characteristic timescale 
that can be controlled by the strength of the coupling. The underlying mechanism is generic, in the sense that 
it depends only on the mixed phase space structure of the underlying uncoupled quasi 1D lattice and may 
therefore apply to a variety of physical systems.

\begin{acknowledgments}
B.L. acknowledges funding  by  a  Marie  Curie  Intra  European Fellowship (G.A.  no  654908) 
within  Horizon  2020 and A.K.M acknowledges a doctoral research grant (Funding ID: 57129429) by the Deutscher Akademischer Austauschdienst (DAAD). T.X. acknowledges financial support by the China Scholarship Council (CSC) during a visit to the University of Hamburg.
\end{acknowledgments}

\bibliographystyle{apsrev4-1}
\bibliography{mybib}

\end{document}